# Anomalous Ultrafast Thermalization of Photoexcited Carriers in Two-Dimensional Materials Induced by Orbital Coupling*

Zhuoqun Wen(文卓群)[1,2,3], Haiyu Zhu(诸海渝)[2,4], Wenhao Liu(刘文浩)[5], Zhi Wang(王崎)[5,**], Wen Xiong(熊稳)[1,3,†], and Xingzhan Wei(魏兴战)[1,2,3,‡]

[1] Chongqing Institute of Green and Intelligent Technology, Chinese Academy of Sciences, Chongqing 400714, China
[2] Chongqing School, University of Chinese Academy of Sciences, Chongqing 400714, China
[3] University of Chinese Academy of Sciences, Beijing 100049, China
[4] School of Optoelectronic Engineering, Chongqing University of Posts and Telecommunications, Chongqing, 400065, China
[5] State Key Laboratory of Semiconductor Physics and Chip Technologies, Institute of Semiconductors, Chinese Academy of Sciences, Beijing 100083, China

*Supported by the Natural Science Foundation of Chongqing of China under Grant No. CSTB2023NSCQ-LZX0087, and the National Natural Science Foundation of China under Grant No.62074021 and No.12174380.

**Corresponding author: wangzhi@semi.ac.cn
†Corresponding author: xiongwen@cigit.ac.cn
‡Corresponding author: weixingzhan@cigit.ac.cn

Understanding the dynamics of photoexcited carriers is essential for advancing photoelectronic device design. Photon absorption generates electron-hole pairs, and subsequent scatterings can induce ultrafast thermalization within a picosecond, forming a quasi-equilibrium distribution with overheated electrons. The high-energy tail of this distribution enables carriers to overcome energy barriers, thereby enhancing quantum efficiency—a phenomenon known as photo-thermionic emission (PTE). Despite its importance, the onset and mechanisms of PTE remain under debate. Using real-time time-dependent density functional theory (rt-TDDFT), we investigate ultrafast carrier thermalization in two-dimensional materials graphene and $PtTe_2$, and the results reveal distinct differences. In graphene, both electrons and holes thermalize into Fermi-Dirac distributions with good agreement to experiment, while $PtTe_2$ exhibits anomalous high-energy tails for both electrons and holes, deviating significantly from Fermi-Dirac behavior. We attribute this anomaly to differences in orbital coupling between the two materials, from which we derive design principles for identifying optimal PTE candidates and, ultimately, improving photodetector performance.

**PACS:** 71.15.Mb (DFT - condensed matter), 87.15.ht (ultrafast dynamics)

---

Enhancing the performance of photodetectors remains one of the central goals in modern photoelectronics. Within the conventional perspective, two primary mechanisms are recognized for photon detection: (1) photon energy exceeding the band gap of a semiconductor, leading to a direct photoelectronic absorption, or (2) photoexcited carriers with energy surpassing the Schottky barrier at the interface of metal-semiconductor junction, generating a photocurrent. The latter mechanism, known as the internal photoemission (IPE) effect, has been pivotal in the development of visible and near-infrared photodetectors[1-3]. The efficiency and spectral response range of photodetectors are hence fundamentally constrained by the Schottky barrier height. However, recent experiments on heterojunction devices of two-dimensional materials, particularly graphene-silicon systems, have challenged this limitation. These systems demonstrated substantial photocurrent responses to long-wavelength infrared radiation, even in the presence of high Schottky barriers[4-8]. The PTE effect has been proposed to explain this discrepancy[9-13]. Being observed in ultrafast transient measurements of graphene[14, 15], PTE describes a non-equilibrium process in which the photoexcited carriers undergo ultrafast thermalization via electron-electron (or hole-hole) scattering, forming quasi-equilibrium distributions, e.g., a Fermi-Dirac distribution, within sub-picoseconds. The high-energy tail of this distribution enables carriers to overcome energy barriers, thereby generating photocurrent and enhancing quantum efficiency. The two-temperature model (TTM), in which electrons can have a high temperature while the lattice remains relatively cool, has been widely adopted to describe such system, and the energy transfer through the electron-phonon coupling gradually restores the thermal equilibrium between electrons and the lattice over a longer timescale[16-18].

While the TTM has been successful in describing photoelectronic processes in graphene-based heterostructures, questions remain about its general applicability. Specifically, is the Fermi-Dirac distribution always valid and "temperature" always well-defined across different systems? More importantly, how and when can the carrier occupation in the high-energy tail be tuned to enhance photodetector efficiency? To address the complexities observed in ultrafast dynamics, extensions to the TTM have been developed[19, 20]. These include non-thermal lattice models[21, 22], three-temperature models[23], and alternative theoretical frameworks such as time-dependent Boltzmann equations[24-29], master equations[30], non-equilibrium Green's functions[31, 32], and TDDFT[33-39]. While conventional models primarily attribute the thermalization to electron-electron scattering[14, 40-43], alternative perspectives highlight the role of phonon interactions[44-47] or orbital coupling[35-37, 48, 49] in shaping carrier dynamics. Nonetheless, the lack of first-principles studies limits a comprehensive understanding of thermalization processes, and the mechanisms underlying the formation of overheated quasi-equilibrium distributions remain contentious.

In this study, we use rt-TDDFT[34] to investigate the ultrafast carrier thermalization dynamics in two-dimensional materials, monolayer graphene and PtTe$_2$. We employ an approach[50] based on band-to-band transition coefficients to examine the time-resolved evolution of carrier distributions, not imposed *a priori* but computed directly from first-principles simulations. In graphene, we demonstrate that photoexcited carriers rapidly achieve a Fermi-Dirac distribution within ultrafast timescales, consistent with experimental observations. In contrast, in PtTe$_2$, we uncover significant deviations from Fermi-Dirac behavior in the distribution of photoexcited carriers, characterized by the presence of overheated electrons and holes. These anomalous thermalization effects are attributed to the distinctive orbital coupling mechanisms in PtTe$_2$ compared to graphene. We then propose design principles for identifying or engineering high-efficiency PTE materials based on orbital coupling and consistency, providing a roadmap for future material development.

**Real-time time-dependent density functional theory.** Our rt-TDDFT approach is designed to solve non-perturbative, non-linear dynamics in systems of hundreds of atoms over timescales ranging from femtoseconds to picoseconds[34, 50-52]. The concepts of this method are elaborated in Ref. [34] and summarized as follows: The time-dependent electron wavefunctions $\{\psi_i(t)\}$, evolving according to the time-dependent Schrödinger equation $i\partial\psi_i(t)/\partial t = H(t)\psi_i(t)$, can be expanded in terms of the adiabatic states $\{\phi_j(t)\}$,

$$\psi_i(t) = \sum_j C_{j,i}(t)\phi_j(t) \quad (1)$$

Here, $\{\phi_j(t)\}$ are the eigenstates of time-dependent Hamiltonian, $H(t)\phi_j(t) = \varepsilon_j(t)\phi_j(t)$. The nuclear positions follow Newton's law, with forces derived from the total energy via Hellmann-Feynman theorem. In our rt-TDDFT implementation, electron-electron interactions are treated within a single-determinant framework as in conventional DFT approaches, i.e., in the Kohn-Sham potential including the exchange-correlation terms.

To analyze the carrier distribution, the time-dependent density of state $DOS(\varepsilon, t)$ and time-dependent density of occupied state $Occ(\varepsilon, t)$ at energy $\varepsilon$ and time $t$ are calculated as follows

$$DOS(\varepsilon, t) = g \sum_j \delta(\varepsilon - \varepsilon_j) \tag{2}$$

$$Occ(\varepsilon, t) = \sum_j \delta(\varepsilon - \varepsilon_j) \sum_i Occ^0(\varepsilon_i^0) |C_{j,i}(t)|^2 \tag{3}$$

where the summation over $j$ runs over all adiabatic states, and the summation over $i$ includes all time-dependent states. $g$ is the spin degeneracy factor, $Occ^0(\varepsilon_i^0)$ is the initial occupation on state $i$ at time zero (before laser), $C_{j,i}(t)$ is the expansion coefficient in Eq.(1), and $\delta(\varepsilon - \varepsilon_j)$ represents the smearing function with a broadening factor. The density of excited carriers at time $t_1$, $D_{exci}(\varepsilon, t_1)$, can then be calculated by the difference between $Occ(\varepsilon, 0)$ and $Occ(\varepsilon, t_1)$, which we mark as

$$D_{exci}(\varepsilon, t_1) = Occ(\varepsilon, t_1) - Occ(\varepsilon, 0) \tag{4}$$

$D_{exci}$ can be either positive or negative, where positive values indicate excited electrons and negative ones indicate excited holes. The time-dependent distribution function of electrons $f(\varepsilon, t)$ is derived from the equation

$$Occ(\varepsilon, t) = f(\varepsilon, t) \times DOS(\varepsilon, t) \tag{5}$$

and the distribution function of excited carrier at time $t_1$, $f_{exci}(\varepsilon, t_1)$, is defined as the difference between $f(\varepsilon, 0)$ and $f(\varepsilon, t_1)$,

$$f_{exci}(\varepsilon, t_1) = f(\varepsilon, t_1) - f(\varepsilon, 0) \tag{6}$$

This approach ensures that the carrier distribution $f$ is not imposed *a priori*, but is instead computed directly from first-principles simulations.

The DFT and rt-TDDFT simulations are implemented in the PWmat package. All runs have been calculated using norm-conserving pseudopotentials (NCPP) and Perdew-Burke-Ernzerhof (PBE) functional[53] (see the Supplementary Materials for details).

**Ground-state properties of graphene and PtTe$_2$.** We first calculated the lattice constants and band structures for monolayer graphene and PtTe$_2$, and compared the results with previous literature. As summarized in Table 1, our calculated lattice constants of graphene and PtTe$_2$ deviate from experimental values[54, 55] by approximately 0.05% and 0.3%, respectively. The ground-state DFT simulations for graphene have been extensively validated in previous studies[56-59]. Therefore, we focus on monolayer PtTe$_2$, and its calculated band structure is presented in Fig. 1. The result reveals an indirect bandgap of 0.34 eV. Both the gap energy and the E-vs-***k*** dispersion agree well with prior studies[55, 60-62].

**Table 1 Comparison of lattice constants of monolayer graphene and PtTe$_2$.**

|  | This work | Previous theory | Experiment |
|---|---|---|---|
| Graphene | 2.458 Å | 2.4589 Å [a]<br>2.4595 Å [b]<br>2.46 Å [c] | 2.4589±0.0005 Å (297 K) [d] |

| | | 4.05 Å (PBE+D3) [e] | |
| PtTe$_2$ | 4.043 Å | 4.1 Å (PBE+TS) [e] | 4.03 Å [e] |
| | | 4.06 Å (PBE+MBD) [e] | |

[a] Ref. [56]  [b] Ref. [57]  [c] Ref. [58]  [d] Ref. [54]  [e] Ref. [55]

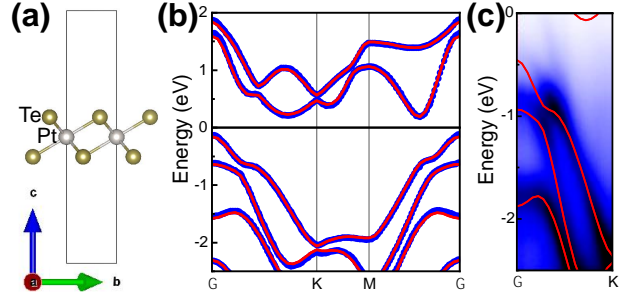

**Fig. 1 Ground-state properties of monolayer PtTe$_2$.** (a) Schematic diagram of the atomic structure. (b) Comparison of band structures between this work (red) and the previous study in Ref. [55] (blue). (c) Dispersive features along the Γ-K direction, showing a comparison between this work (red solid lines) and experimental ARPES data from Ref. [55] (black-blue-white colormap).

**Photoexcited carrier thermalization in graphene.** Using a laser field similar to the one reported in experiment Ref. [15] (See the Supplementary Materials for details in the "Computational Parameters" section), we investigate the photoexcited carrier distributions $f_{\text{exci}}(\varepsilon,t)$ obtained from TDDFT and compare them with the ones extracted from energy distribution curves reported in the same reference. Fig. 2 illustrates the simulated distribution functions of the excited carriers from TDDFT. Both electrons and holes thermalize into well-defined Fermi-Dirac distributions, with goodness of fit $R^2 > 0.9$. Such quasi-equilibrium distributions are formed within 30 fs (Fig. 2(b)), a timescale aligns well with that from ultrafast experiment but significantly shorter than that typically required for electron-phonon coupling. The temperature of the excited electrons after 30 fs, extracted from the Fermi-Dirac fit, is approximately 3000 K, which is in good agreement with experimental results in Ref. [15]. The Fermi-Dirac behavior is maintained throughout the simulation (to 240 fs, Fig. 2(c)). while there is no evident cooling or recombination processes observed, suggesting that these phenomena may occur over longer timescales.

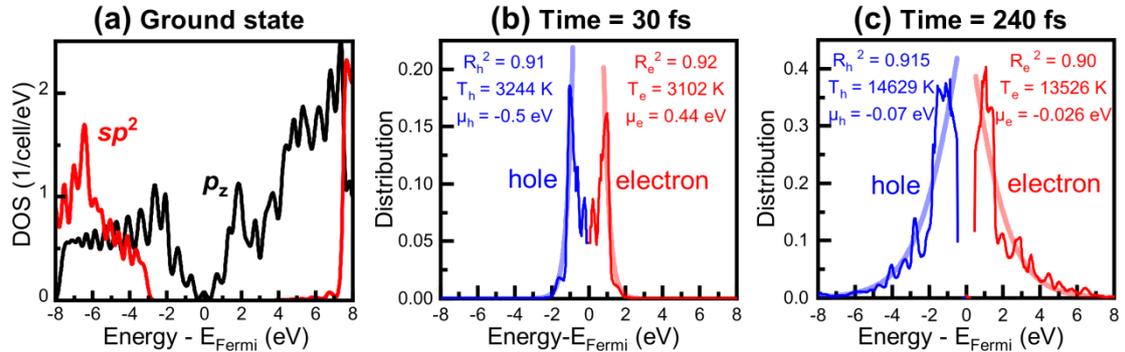

**Fig. 2 Carrier distributions in monolayer graphene.** (a) The calculated ground-state density of states of monolayer graphene, projected onto the in-plane, $sp^2$ hybrid orbital (red) and the out-of-plane $p_z$ orbital (black). (b)(c) The distribution functions of photoexcited electrons (red, thin line) and holes (blue, thin line) after 30 fs (b) and 240 fs (c). Note that for holes the distribution is plotted by the absolute values (reversed in sign) to provide a direct comparison

to electrons. Shaded, thicker lines represent the corresponding Fermi-Dirac fits. The fitting parameters, including the goodness of fit ($R_e^2$,$R_h^2$), temperature ($T_e$,$T_h$), and chemical potential ($\mu_e$,$\mu_h$), are labeled next to the fitted curves.

**Photoexcited carrier thermalization in PtTe$_2$.** The evolution of the carrier distribution in monolayer PtTe$_2$, however, exhibits significant differences compared to graphene. Fig. 3 illustrates $f_{\text{exci}}(\varepsilon,t)$ at several different time steps for monolayer PtTe$_2$ exposed to a 1.5 eV laser. The dashed vertical lines mark the region from -1.5 eV to 1.5 eV (setting the valence band maximum as the energy zero), corresponding to the single-photon absorption region. Beyond this region lie the overheated electron region (>1.5 eV) and overheated hole region (<-1.5 eV), as shown in Fig. 3(a).

At the initial stage of laser incidence (30 fs, Fig. 3(a)), almost all excitations occur within the single-photon absorption region. At this point, the excited electrons have already thermalized to a Fermi-Dirac distribution with a tail extending to higher energies, while the holes do not follow the Fermi-Dirac distribution. After that, the excited carriers quickly diverge from the Fermi-Dirac function in both the low- and high-energy regions. Specifically: (1) For the electron, an internal gap develops between 2 eV and 2.5 eV, with excited electrons accumulating around 2.6 eV, forming a small peak (Figs. 3(b)-3(d)). This results in a non-negligible larger number of overheated electrons than what would be expected from a typical Fermi-Dirac high-energy tail. (2) For the hole, it exhibits a broad and continuous distribution extending from -1.6 eV to -2.8 eV, indicating a remarkably high population of overheated holes. These anomalies persist throughout the simulation period, with both overheated electrons and holes maintaining their presence.

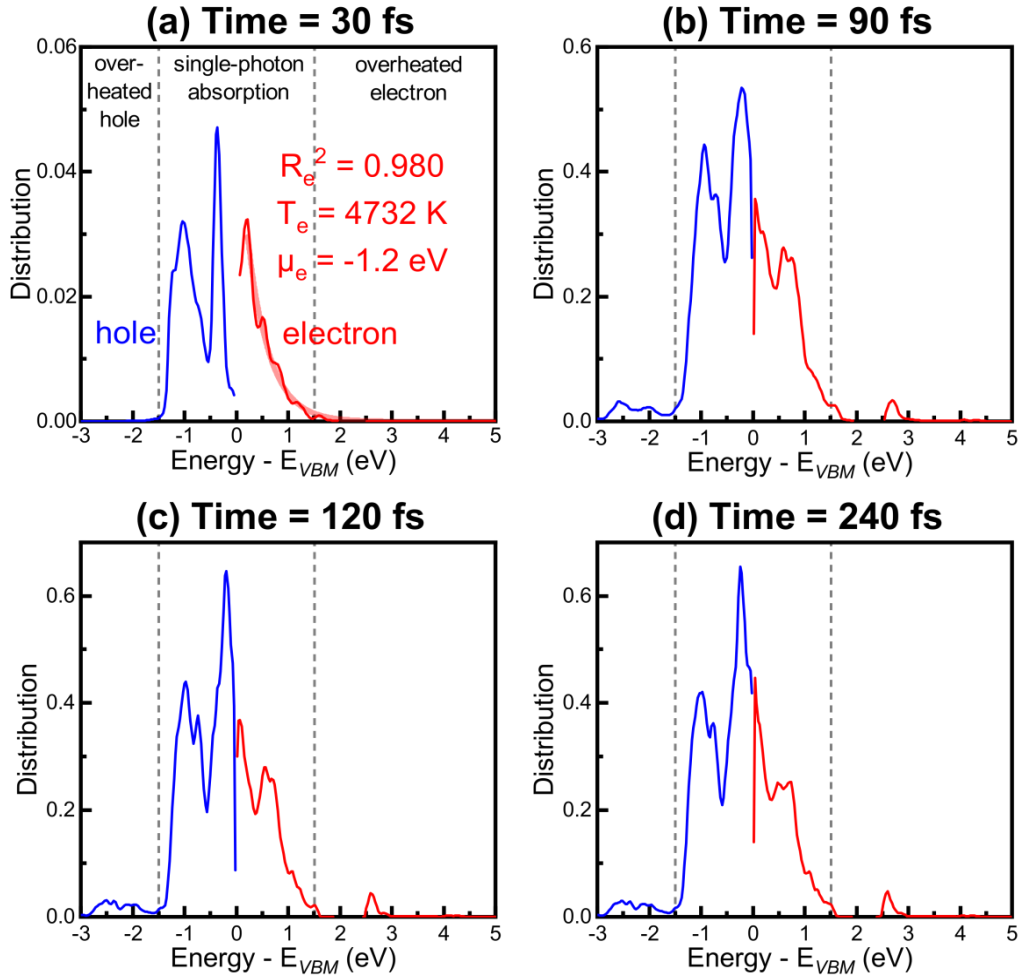

**Fig. 3 The distribution functions of photoexcited electrons (red line) and holes (blue line) in monolayer PtTe$_2$.** The two vertical dashed lines separate the three regions, i.e., the single-photon absorption, overheated electrons, and overheated holes. Note that only the electron distribution in (a) exhibits an obvious Fermi-Dirac behavior, where the Fermi-Dirac fit is shown by a shaded, thicker red line. No other distribution exhibits significant goodness of Fermi-Dirac fit.

Indeed, distribution function $f_{\text{exci}}(\varepsilon, t)$ alone is insufficient to fully describe the photoresponse. In fact, the photocurrent is expected to be proportional to the total number of overheated carriers, which directly correlates with the density of excited carriers $D_{\text{exci}}(\varepsilon, t)$, the product of the distribution function $f_{\text{exci}}(\varepsilon, t)$ and the density of states $\text{DOS}(\varepsilon, t)$, as derived by Eqs.(4)-(6). To understand the non-trivial thermalization, it is necessary to analyze the detailed dynamics of total number of excited carriers and the $D_{\text{exci}}(\varepsilon, t)$. Fig. 4(a) shows the percentage of excited electrons relative to the total valence electrons in PtTe$_2$ over time. Under the influence of the oscillating electric field with a Gaussian envelope, electrons are gradually excited from the valence to conduction bands. At the end of the laser (approximately 120 fs), the excitation stabilizes, with approximately 1.3% of the valence electrons transitioning to the conduction band. While, to the best of our knowledge, no direct experimental data has been reported for PtTe$_2$ on the number of ultrafast photoexcited electrons, this value of ~1% is comparable to results for similar compounds reported in previous studies[63, 64]. Figs. 4(b)-4(e) show the density of excited carriers $D_{\text{exci}}(\varepsilon, t)$ at 30, 90, 120, and 240 fs, respectively. Positive values represent excited electrons while negative values correspond to excited holes. At t=30 fs (Fig. 4(b)), the laser field intensity remains low, and the total energy input from laser is relatively small. As a consequence, there are only limited numbers of excited electrons and holes. The photoexcitation occurs predominantly in the single-photon absorption region, indicating energy-conserving, single-particle excitation processes. As the laser intensity increases, a significant number of electron-hole pairs are generated, and $D_{\text{exci}}(\varepsilon,t)$ demonstrates non-monotonic behavior at 2.6 eV for electrons and from -1.6 eV to -2.8 eV for holes, the same energy regions as observed in $f_{\text{exci}}(\varepsilon,t)$ in Fig. 3. This correlation indicates that the anomalies originate from the excitation processes themselves rather than being an artifact of the density of states. Notably, the number of overheated holes is significantly larger than that of overheated electrons. Such asymmetry between electron and hole is absent in graphene, indicating an asymmetric PTE efficiency for PtTe$_2$ between n-type and p-type heterojunctions.

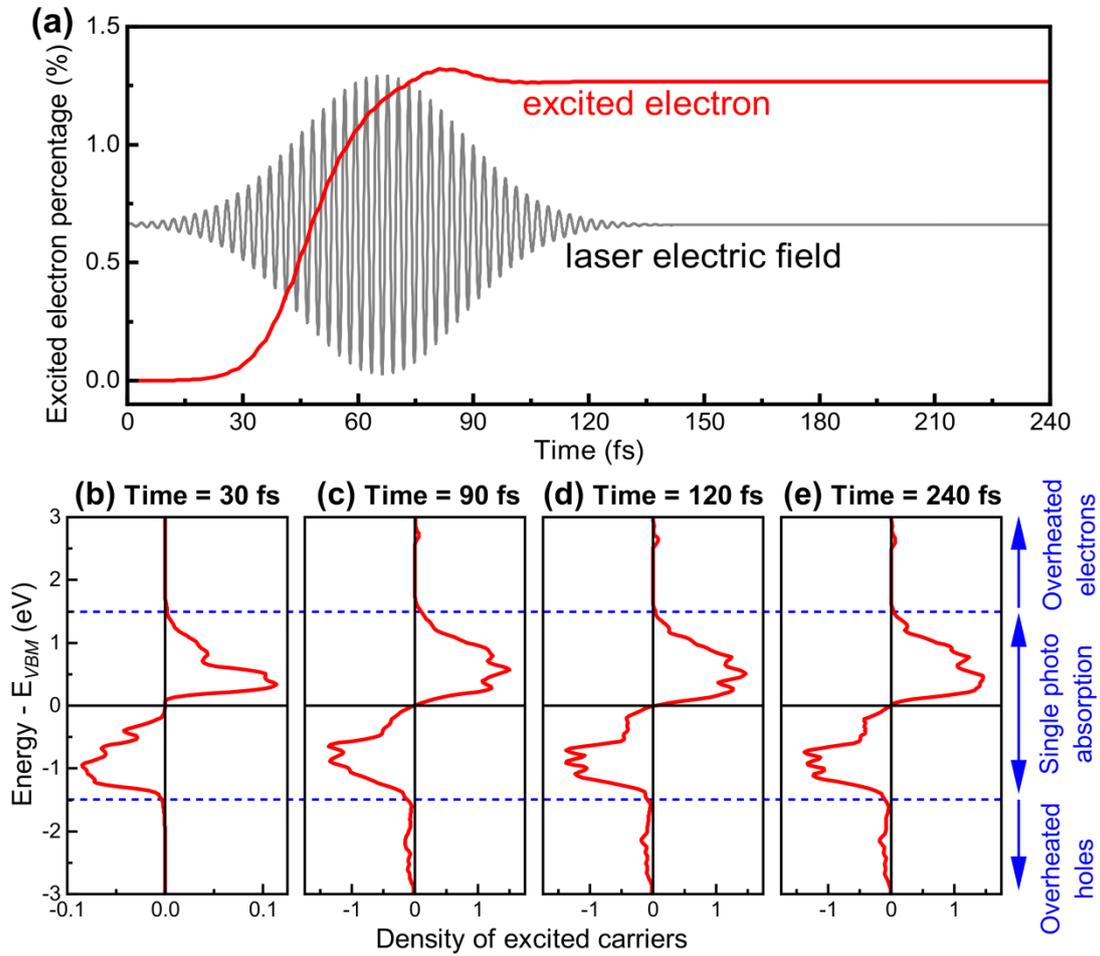

**Fig. 4 The time-dependent total number of excited electrons and density of excited electrons in PtTe$_2$, from TDDFT calculations.** (a) Total number of excited electrons (red, shown as the percentage relative to all valence electrons) and laser electric field (gray) over time. (b-e) Density of excited carriers including electrons (positive values) and holes (negative values) at (b) 30 fs, (c) 90 fs, (d) 120 fs, (e) 240 fs.

**The crystal-field splitting and orbital coupling in PtTe$_2$.** Now, we discuss the physical mechanism behind such anomalous thermalization. In PtTe$_2$, the Pt ions are positioned at the center of distorted (PtTe$_6$) octahedra with D$_{3d}$ site symmetry. The schematic of Fig. 5(a) summarizes the hierarchy of crystal-field splitting that lifts the degeneracy of electron states. In a non-distorted octahedron with O$_h$ symmetry, the 5$d$ orbitals of Pt split into $t_{2g}$ and $e_g$ degeneracies. The D$_{3d}$ distortion induces further splitting among the $e_g$, leading to $e_g''$, $a_{1g}$, and $e_g'$ states. The Pt$^{4+}$ ions have a valence electron configuration of 5$d^6$, resulting in the energy gap between $a_{1g}$ (VBM) and $e_g''$ (CBM) states, marked by light blue in Fig. 5(a). Once combined with the ligand configuration of Te$^{2-}$, the complete orbital configuration near the Fermi level can be described, as shown in Fig. 5(b). The VBM in single-photon absorption region and deeper VB in the overheated hole region consist of $d_{xy}$, $d_{yz}$, and $d_{xz}$ orbitals from Pt 5$d$, bonding with the π orbitals from Te 5$p$ ($p_\pi$); the CB in single-photon absorption region has $d_{x2-y2}$ and $d_{z2}$ from Pt, and the σ orbitals from Te 5$p$ ($p_\sigma$). Above 2.5 eV (overheated electron region) there exists the $s$-$p$ coupling which is mainly Pt 6$s$ and Te 5$p$ ($p_\sigma$). The orbital configuration in PtTe$_2$ is in distinct differences to that of graphene, the latter of which (Fig. 2(a)) is dominated by a single orbital ($p_z$) in a large energy region near the Fermi level.

After the laser is turned on, the single-particle optical transitions predominantly occur between the $a_{1g}$→$e_g''$ states, as governed by selection rules and the fact that the VBM is mainly Te 5$p$ orbitals. Consequently, the initially excited electrons and holes are expected to predominantly occupy Te 5$p$ and Pt 6$d$ orbitals, respectively. Now, the differences in thermalization efficiency between electrons and holes can be understood through the distinct orbital coupling mechanisms they follow:

(1) Electrons. Electrons in the low-energy states can thermalize to higher energies via two possible routes. (i) Pt 6$d$ → Pt 6$s$. This route is inefficient due to the weak orbital coupling between 6$d$ and 6$s$ states. (ii) Pt 6$d$ → Te 5$p$ → Pt 6$s$. This route involves a two-step process with charge transfer between the ion (Pt) and the ligand (Te), and the efficiency is hence limited. Orbital-projected densities of excited electrons (Figs. 5(c)–5(f)) confirm that the high-energy electrons observed at the 2.6 eV peak reside in the Pt 6$s$ orbital, supporting the explanation above.

(2) Holes. Holes exhibit higher thermalization efficiency because they do not require charge transfer between orbitals for energy redistribution. The low- and high-energy states share the same orbital configuration, primarily derived from Te 5$p$ orbitals. This orbital consistency enables smoother energy transitions and more efficient thermalization.

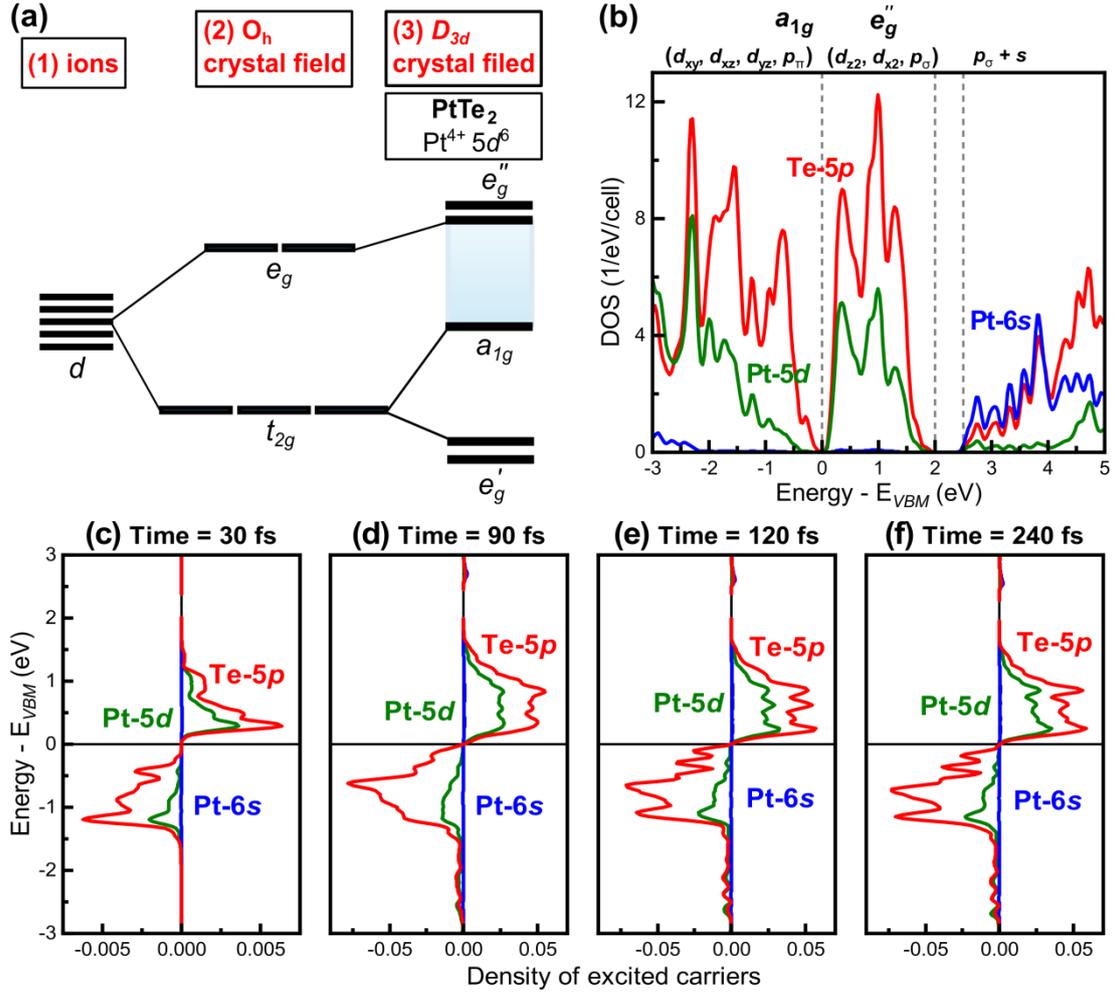

**Fig. 5** (a) Schematic of crystal-field splitting in PtTe$_6$ cluster. (b) Ground-state orbital-projected DOS of PtTe$_2$. (c-f) Orbital-resolved density of excited carriers in PtTe$_2$ at (c) 30 fs, (d) 90 fs, (e) 120 fs and (f) 240 fs.

**Design principles for optimal PTE materials.** Based on the orbital coupling mechanism discussed above, we propose the following design principles for identifying or engineering compounds with high PTE efficiency. A necessary (but not sufficient) condition for achieving high PTE efficiency is the presence of favorable orbital dynamics that facilitate energy transfer or redistribution. Specifically:

(1) *Strong orbital coupling across energy regions.* High PTE efficiency can be achieved if there is significant coupling between the primary orbitals in the single-photon absorption region (low-energy states) and those in the target high-energy region. Such coupling ensures efficient energy redistribution and minimizes energy loss during carrier thermalization.

(2) *Orbital consistency across energy regions.* Alternatively, if the low- and high-energy regions share the same dominant orbitals, energy redistribution becomes seamless, as no charge transfer or interorbital transitions are required. This condition reduces thermalization bottlenecks, ensuring rapid carrier relaxation and high PTE efficiency.

Such design principles can be used to identify candidate materials, or to apply modifications on the orbital compositions of VBM and CBM in materials lacking inherent orbital coupling or consistency, e.g., doping, strain engineering, or heterostructuring.

**Summary.** In this study, we employed rt-TDDFT to investigate the ultrafast thermalization dynamics of photoexcited carriers in monolayer graphene and PtTe$_2$, revealing the fundamental mechanisms driving PTE and outlining design principles for optimizing material performance. In graphene, carriers rapidly thermalize into Fermi-Dirac distributions within ultrafast timescales, consistent with experimental results and indicative of efficient energy redistribution. In PtTe$_2$, however, carrier distributions exhibit marked deviations from Fermi-Dirac behavior, characterized by persistent high-energy tails resulting from unique orbital coupling mechanisms. These findings establish the critical role of orbital interactions in shaping ultrafast dynamics, and provide design principles for identifying or engineering high-efficiency PTE materials based on orbital coupling and consistency. This study advances the understanding of ultrafast carrier dynamics, highlights the limitations of conventional thermalization models, and underscores the need for material-specific considerations in designing next-generation photoelectronic devices.